\documentclass[%
 reprint,
superscriptaddress,
amsmath,amssymb,
prl
]{revtex4-1}

\usepackage{graphicx}
\usepackage{dcolumn}
\usepackage{bm}
\usepackage{xcolor}
\definecolor{darkblue}{rgb}{0,0,.5}
\definecolor{darkgreen}{rgb}{0,.5,0}
\definecolor{darkestred}{rgb}{.7,0,0}
\definecolor{darkred}{rgb}{0.9,0,0}
\usepackage{tikz}

\begin{document}
\preprint{APS/123-QED}

\title{On the coupling between buoyancy forces and electroconvective instability near ion-selective surfaces }

\author{Elif Karatay}
\email{karataye@gmail.com}
 \affiliation{
 Department of Mechanical Engineering, Stanford University and Center for Turbulence Research, Stanford University, Stanford, California 94305, USA \\
}
\author{Matthias Wessling}
\affiliation{
 RWTH Aachen University, Aachener Verfahrenstechnik, 52056 Aachen, Germany\\
}
\author{Ali Mani}
\email{alimani@stanford.edu}
\affiliation{
 Department of Mechanical Engineering, Stanford University and Center for Turbulence Research, Stanford University, Stanford, California 94305, USA \\ 
}

\date{\today}

\begin{abstract}
Recent investigations have revealed that ion transport from aqueous electrolytes to ion-selective surfaces is subject to electroconvective instability that stems from coupling of hydrodynamics with electrostatic forces. Electroconvection is shown to enhance ion mixing and the net rate of transport. However, systems subject to electroconvection inherently involve fluid density variation set by salinity gradient in the bulk fluid. In this study we thoroughly examine the interplay of gravitational convection and chaotic electroconvection. Our results reveal that buoyant forces can significantly influence the transport rates, otherwise set by electroconvection, when the Rayleigh number $Ra$ of the system exceeds a value $Ra \sim 1000$. We show that buoyancy forces can significantly alter the flow patterns in these systems. When the buoyancy acts in the stabilizing direction, it limits the extent of penetration of electroconvection, but without eliminating it. When the buoyancy destabilizes the flow, it alters the electroconvective patterns by introducing upward and downward fingers of respectively light and heavy fluid. 
\end{abstract}

\maketitle
Mass transfer beyond diffusion limitation is possible in electrically driven systems when ions are transported from a fluid electrolyte to a charge selective interface \textit{e.g.} an ion exchange membrane or an electrode. Among several mechanisms on the origins of this \textit{over-limiting} mass transfer, electroconvection, which plays a crucial role in ion mixing, has been suggested as one of the key mechanisms.~\cite{Nam} In such systems when the applied voltage is above a threshold (for example $\gtrsim 0.5~V$ for aqueous systems with monovalent ions at room temperature), oscillations in the instantaneous current signal have been measured while an over-limiting current is sustained.~\cite{Lifson1978, Krol} The noise in the electric response has been attributed to induced convection~\cite{Lifson1978, Rubinstein1988, Krol, Zabolotsky1998, Pismenkaya2004} and indeed flow vortices have been observed in experiments~\cite{Lammertink, Chang}. Consistent with these observations, a theoretical analysis has shown that ion transport across charge-selective interfaces is prone to electrokinetic instabilities (EKI) stemming from a coupling of the fluid flow with ion-transport and electrostatic interactions~\cite{Zaltzman2007}. More recent direct numerical simulations demonstrated transitions from regular coherent vortices to chaotic multi-scale structures when the applied potential is higher than a limit ($\gtrsim 1V$)~\cite{Druzgalski, Karatay,Davidson}. In all these studies, however, gravitational effects have been neglected. 

\begin{figure} [h!]
\includegraphics[width=8.1cm]{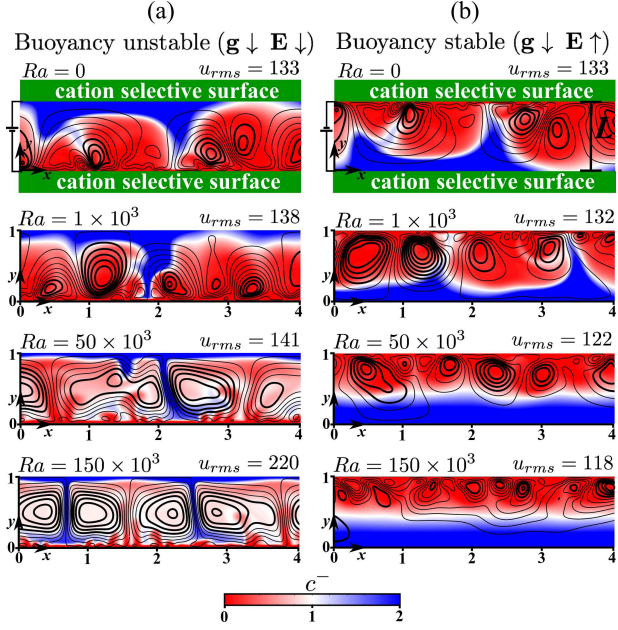}
\caption{\label{fig:fig1} Aqueous monovalent salt solution between two horizontal cation selective surfaces subject to DC electric field in a gravitationally unstable (a) and stable arrangement (b). Surface plots show the fully developed anion concentration fields at $t=0.2$ superimposed with flow streamlines. 
Here $\Delta \phi=80V_T$, $\kappa=0.5$. The $u_{rms}$ is scaled by $D/L$. }
\end{figure}

Systems prone to EKI experience fluid density gradients even before the onset of EKI due to strong salt concentration gradients associated with ion-concentration polarization phenomenon~\cite{Zaltzman2007, Nikonenko2014}. Previous studies have considered gravitational effects in electrochemical systems, but ignored electrokinetic effects. The requirements for the onset of gravitational fluid instabilities are often satisfied when the Rayleigh number $Ra=\beta \Delta c g L^3/ \nu D$ of the system exceeds a critical value in electrolytic systems~\cite{Nikonenko2014, Maletzki, Deng, Lifson1978, Zabolotsky1996, Zabolotsky1998, Rubinstein1988, Krol, Wilke, Fenech1960, Hage, Ward, Gonzalez2002, Pismenkaya2004, Pismenkaya2006}. Here $L$ is the distance from bottom to top boundary; $g$ is the magnitude of the gravitational acceleration; $\beta$ and $D$ are the solute expansion coefficient and diffusivity of the 
electrolyte; $\nu$ is the kinematic viscosity of the fluid; and $\Delta c$ represents the scale for the variation in the concentration. Provided that $Ra$ number is sufficiently large in a gravitationally unstable arrangement (Fig.~\ref{fig:fig1}a), flow instabilities arise due to the action of buoyant forces. Buoyant flow structures have been experimentally observed near ion selective interfaces~\cite{Hage, Marshall2006, Muhlenhoff, Bruyn}. Their significant impact on the limiting current $i_{lim}$~\cite{Volgin, Rubinstein1988, Fenech1960, Wilke, Zabolotsky1998, Zabolotsky1996, Goldstein} and the morphology of the electrodeposits~\cite{Bruyn, Huth, Marshall2006, Soba} have been reported in low voltage range experiments for decades. In fact electrochemical experiments have been also used to model the buoyancy driven flow in the Rayleigh-Benard (RB) convection problem.~\cite{Goldstein, Hage, Fenech1960} Most of these experiments were set at voltages close to the onset of limiting current to achieve $c \approx 0$ close to the ion-absorbing boundary and thus ensure a well defined Rayleigh number.~\cite{Goldstein, Winkler1996, Hage, Ward} This regime happens to be well-below the threshold for EKI effects. However, most practical scenarios such as \cite{Huth} and \cite{Gonzalez2002, Marshall2006, Mocskos2011} involve regimes where the applied voltage and system Rayleigh number are both above their critical values and thus inevitably involve coupling of electroconvection with buoyancy effects.  

Despite decades of research revealing the significance of gravitational convection as well as electrokinetic instabilities in electrochemical systems, a quantitative understanding of coupling between these effects is not yet developed. In this Letter, we thoroughly examine the interplay of gravitational convection and chaotic electroconvection via 2D direct numerical simulations. 

We consider an aqueous solution of a univalent salt in between two horizontal cation selective surfaces at $y=0$ and $y=1$. (Fig.~\ref{fig:fig1}) The fluid electrolyte is initially at a quiescent state with a uniform bulk concentration $c_{b}$. Incompressible Navier-Stokes and Poisson-Nernst-Planck equations describe the flow, electric potential and ion transport respectively. 
\begin{subequations}
\label{eq:modeleqns}
\begin{eqnarray}
\frac{1}{Sc} \left[ \frac{\partial \bm{u}}{\partial t}+\bm{u}.\nabla \bm{u} \right]=-\nabla p'+\nabla^2\bm{u}+\bm{f_e} + \bm{f_g}, \label{subeq:NS}
\end{eqnarray}
\begin{equation}
\nabla.\bm{u} =0, \label{subeq:continuity}
\end{equation}
\begin{equation}
-2\epsilon^2 \mathbf{\nabla}^2 \phi=\rho_e,\label{subeq:Poisson}
\end{equation}
\begin{equation}
\frac{\partial c^{\pm}}{\partial t}=-\nabla\bm{ j^{\pm}},\label{subeq:NP}
\end{equation}
\end{subequations}
Here $\bm{u}=u\bm{\hat{x}}+ v\bm{\hat{y}}$ is the velocity vector field; $p'$ is the modified pressure including the hydrostatic effects; $c^{+}$ is the cation concentration; $c^{-}$ is the anion concentration; $\phi$ is the electric potential; $\rho_{e} = z(c^+ - c^-)$ is the free charge density with ionic valence $z=\pm 1$. 
$\bm{f_e} = -\kappa \rho_{e}\bm{\nabla}\phi/2\epsilon^2$ is the electrostatic body force and $\bm{f_g}=-\bm{\hat{y}}cRa$ is the gravitational body force where $c=(c^{+} + c^{-})/2$ is the salt concentration. $\bm{j^{\pm}} =c^{\pm}\bm{u}-\bm{\nabla}c^{\pm} \mp c^{\pm}\bm{\nabla}\phi$ are ion fluxes of anions and cations. Eqs.~\ref{subeq:NS}-\ref{subeq:NP} are dimensionless where the spatial coordinates, velocity, time, concentrations and electric potential are respectively scaled by domain height $L$, diffusion velocity $D/L$, diffusion time $L^2/D$, bulk concentration $c_{b}$ and thermal voltage $V_T = k_B T/ze$ where $k_B$ is Boltzmann constant and $e$ is elementary charge.

In the Navier-Stokes equations (Eq.~\ref{subeq:NS}), we neglect the nonlinear terms $\bm{u}.\nabla \bm{u}$~\cite{Druzgalski} and we employ the Oberbeck-Boussinesq (OB) approximation~\cite{Koschmieder} in which the fluid density $\rho$ is assumed to be a linear function of the salt concentration $c$; $\rho (c) = \rho_{o}\left[ 1+\beta(c-c_{b})\right]$ where  $\rho_{o}$ is the density of the fluid when $c=c_{b}$. Within the OB approximation, the key dimensionless control parameters of our model problem are the (\textit{i}) Rayleigh number $Ra=\beta c_b g L^3/ \nu D$ (\textit{ii}) the electrohydrodynamic coupling constant $\kappa = \varepsilon V_T^2 /(\nu \rho_{o} D)$ where $\varepsilon$ is the dielectric permittivity, (\textit{iii}) the dimensionless Debye length $\epsilon = \lambda_D / L$ where $\lambda_D=\sqrt{\varepsilon k_B T / [2(ze)^2 c_{b}]}$ is the dimensional Debye length, and (\textit{iv}) the dimensionless applied potential $\Delta \phi$ in units of thermal voltage. In our nondimensionalization, the $Ra$ number is defined in terms of the initial bulk salt concentration $c_b$, instead of $\Delta c$, since in the overlimiting regime, the scale for the variation in the concentration $\Delta c$ is  $\mathcal{O} (c_b)$. We note that this is unlike the thermal RB problem where $\Delta T$ is the controlling parameter. 
\begin{figure}
\centering
  \includegraphics[width=8.7cm]{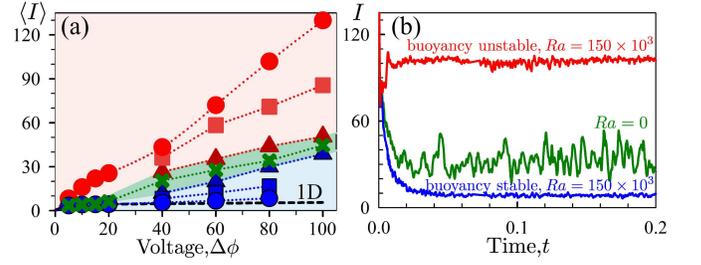}
  \caption{Effect of magnitude and direction of the gravitational force. (a) Time averaged current density $\langle I \rangle$ where the cross symbols $\color{darkgreen}\chi$ represent $Ra=0$ and the green shaded area depict when gravitational effects are negligible. Red shaded area and the blue shaded area depict the gravitationally unstable and stable regime, respectively. Triangles $\blacktriangle$, squares $\blacksquare$ and circles $\bullet$ represent $Ra=1\times 10^3$, $50\times 10^3$ and $150\times 10^3$, respectively. The black dashed line $\linethickness{0.3mm} \line(1,0){7} \, \line(1,0){7}$ presents the one-dimensional $\langle I \rangle$. (b) Instantaneous current density $I$ at $Ra=0$ (green line) and $Ra=150\times 10^3$ for gravitationally unstable (red line) and stable (blue line) configurations. Here $\Delta \phi =80V_T$. In (a)-(b), $\kappa=0.5$}
  \label{fig:fig2}
\end{figure} 
Eqs.~\ref{subeq:NS}-\ref{subeq:NP} are solved in a 2D domain, where periodic boundary conditions are implemented in the $x$-direction over an aspect ratio of 6. No-slip boundary conditions are implemented along the cation selective surfaces at $y=0$ and $y=1$. Zero-flux conditions are enforced for anions $\bm{j^-}=0$ and dimensionless fixed cation concentration $c^+ = 2$ at $y=0$ and $y=1$. The gravity vector $\mathbf{g}$ points in the negative y-direction. To explore both buoyantly unstable and stable systems we consider different cases with difference in downward versus upward direction of the applied electric field. (Fig.~\ref{fig:fig1}) We solve the governing equations with second order finite differences as explained in detail in Ref.~\cite{Karatay}. We studied the dynamics of interplaying buoyancy and electrokinetic effects in experimentally relevant regimes corresponding to a wide range of $Ra$ number ($0\leqslant Ra \leqslant 1 \times 10^{7}$), applied potential $\Delta \phi$ ($5V_T\leqslant \Delta \phi \leqslant 100V_T$) and electrohydrodynamic coupling constant $\kappa$ ($0 \leqslant \kappa \leqslant 0.5$). 

Fig.~\ref{fig:fig1} shows the fully developed anion concentration $c^-$ fields superimposed with the flow lines for varying $Ra$ numbers in gravitationally unstable and stable orientations at a fixed $\Delta \phi=80V_T$. The top panels of Fig.~\ref{fig:fig1}a-b do not include the buoyancy effects ($Ra=0$) and are identical except for the direction of the electric field $\bm{E}$. When the buoyancy effects are considered at $Ra=1\times 10^3$, the root-mean-square of the velocity $u_{rms}$ varies slightly for the gravitationally unstable and stable arrangements although the concentration fields remain qualitatively similar. The significance of the value of $Ra=1\times 10^3$ is that it is on the order of the critical Rayleigh number $Ra_{cr}$ in systems that are governed solely by buoyancy effects~\cite{Baranowski, Volgin, Ward, Koschmieder}. However we demonstrate that for a coupled system, the EKI effects are dominant for $Ra \sim \mathcal{O}(10^3)$ when the applied voltage is sufficiently high. In dimensional units $Ra=1\times 10^3$ corresponds to $c_b \sim$ 1mM for a dilute NaCl solution with a domain length scale of $L \gtrsim$ 1.4 mm.  

When $Ra\gtrsim 1\times 10^3$, the ion distribution and flow fields for buoyantly unstable and stable configurations are tremendously different. In the unstable configuration (Fig.~\ref{fig:fig1}a), RB structures are clearly observed in the form of large scale plumes ($\sim L$) detaching from the boundary layers at the cation selective surfaces. For a moderate $Ra$, \textit{e.g.}~$50\times 10^3$, the electrokinetic chaos prevail even in the large scale RB plumes whereas for a higher $Ra$, \textit{e.g.}~$150\times 10^3$, the large scale plumes become regularized in the form of convection cells. Nevertheless, the small scale EKI vortices are sustained for all Ra (third and fourth panels of Fig 1a) even up to  $Ra \sim \mathcal{O}(10^6)$. This can be vividly seen by a comparison between top (EKI stable) and bottom (EKI unstable) membranes. Our investigations of a wide range of $Ra$ numbers suggest that increasing $Ra$ results in thinner, faster and nearly regular cells consistent with the classical RB flows~\cite{Koschmieder}. 

Fig.~\ref{fig:fig1}b demonstrates gravitationally stable scenarios in which the ion depleted layer is established near the top boundary. In this case, EKI vortices emerge at the upper membrane boundary. The domain partitions into two layers; in the lower layer the EKI vortices are completely suppressed by the buoyancy effects and the transport is dominated by diffusion and electromigration. In the upper layer, while salt is highly depleted, there are still strongly active EKI vortices.  In other words, buoyancy effects place an `edge' on the depth of penetration of EKI vortices. As $Ra$ number is increased, however, the edge of the upper layer is only slightly pushed back. Here the mean salt concentration gradient is very small and associated with the so-called extended space charge layer~\cite{Zaltzman2007}. The $\Delta c$ over this zone, and thus the effective local $Ra$ number, is much smaller than that of nominal $Ra$ number; thereby the buoyancy forces cannot overcome the strong electrostatic body force due to EKI in this upper layer. In this case the relative change in root-mean-square velocity $u_{rms}/u_{rms|_{Ra=0}}$ remains $\approx \mathcal{O}(1)$. Whereas in the gravitationally unstable configuration $u_{rms}/u_{rms|_{Ra=0}}$ can be as high as $\approx 15$. We provide the plots of velocity, $u_{rms}$, free charge, energy and concentration spectra, as well as movies in the Supplemental Material~\cite{SI}. 
\begin{figure} 
\centering
  \includegraphics[width=8.7cm]{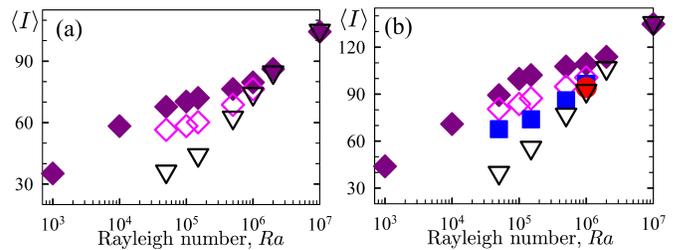}
  \caption{Time averaged current density with respect to $Ra$ number obtained for gravitationally unstable configuration at (a) $\Delta \phi=60V_T$ and (b) $\Delta \phi =80V_T$ for various $\kappa$. In (a)-(b) $\kappa=0.5$ ($\color{violet}\blacklozenge$), $\kappa=0.2$ ($\color{magenta}\diamondsuit$), $\kappa=0.1$ ($\color{blue} \blacksquare$), $\kappa=0.05$ ($\color{red} \bullet$) and $\kappa=0$ ($\triangledown$). }
\label{fig:fig3}
\end{figure} 

\begin{figure*}
\centering
 \includegraphics[width=14.5cm]{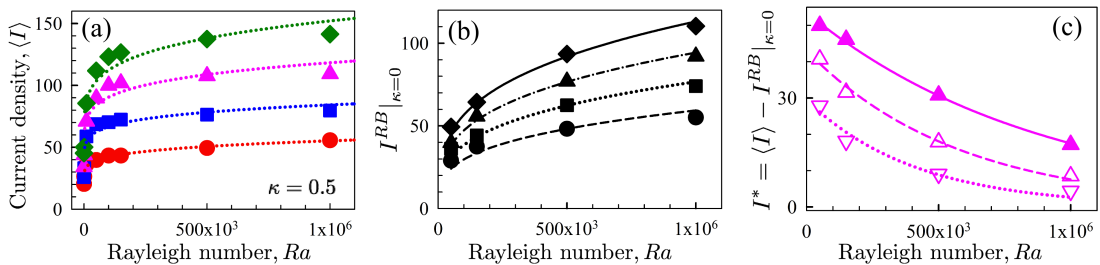}
  \caption{Contributions of RB and EKI induced convection to the current density in a gravitationally unstable arrangement. (a) $\langle I \rangle$  data (symbols) plotted with a power law scaling $\langle I \rangle \sim y_0 + mRa^{n}$ (dotted lines) at a finite $\kappa=0.5$ for various $\Delta \phi$, revealing poor correlations. (b) Current density data obtained at $\kappa=0$ plotted with a power law scaling $\langle I \rangle (\kappa=0)  \sim Ra^{0.3}$ for various $\Delta \phi$. In (a-b) $40V_T$: circles $\bullet$, $60V_T$: squares $\blacksquare$, $80V_T$: triangles $\blacktriangle$ and $100V_T$: diamonds $\blacklozenge$. (c) Current density $I^*$ suggested to be induced by the chaotic EKI vortices with respect to $Ra$ at $\Delta \phi=80V_T$. The symbols present the DNS data obtained at $\kappa=0.5$ ($\color{magenta}\blacktriangle$), $\kappa=0.2$ ($\color{magenta}\triangle$) and $\kappa=0.1$ ($\color{magenta} \triangledown$). Lines present exponential decay fits with respect to $Ra$; $I^* \sim e^{-bRa}$ where $b \sim \mathcal{O}(10^{-6})$ for $\kappa=0.5$ (solid line), $\kappa=0.2$ (dashed line), $\kappa=0.1$ (dotted line).}
  \label{fig:fig4}
\end{figure*} 
Next we quantify the net transport rate in these systems by presenting the time and area averaged current density, $\langle I \rangle$. In a gravitationally unstable arrangement when $Ra \geqslant 1 \times 10^3$, the current density is higher for all $\Delta \phi$ compared to that for $Ra=0$~(Fig.~\ref{fig:fig2}a). In addition, the characteristic plateau region, representing the diffusion limited transport, shortens for higher $Ra$ number.~(Fig.~\ref{fig:fig2}a) These observations are in a very good qualitative agreement with previous experimental reports~\cite{Krol, Lifson1978, Maletzki, Pismenkaya2004, Rubinstein1988, Zabolotsky1996}. The unstable RB plumes enhance the ion mass transfer by enhancing the fluid mixing. Whereas in a gravitationally stabilized position, $\langle I \rangle$ diminishes and approaches to a one-dimensional $\langle I \rangle$ sustained by only electromigration and diffusion. In this case, while EKI vortices are sustained in the upper layer, they do not substantially contribute to the net transport because they only cause mixing within a highly salt-depleted region. 

In~Fig.~\ref{fig:fig2}b, we compare the instantaneous current $I$ obtained at $Ra=0$ and $Ra=150 \times 10^3$ for the gravitationally unstable and stable configurations. The buoyancy forces damp the oscillations in $I$ by regularizing the flow in both RB stable and unstable conditions. This observation also explains previous measurements in electrochemical cells~\cite{Ward, Lifson1978}. The amplitude of the oscillations are lower in the gravitationally stable position whereas the oscillation period is similar for both configurations, determined by the frequency of EKI induced vortices thus by $\Delta \phi $ and $\kappa$ at a given $Ra$ number. 

Our current density $\langle I \rangle$ results plotted with respect to $Ra$ number (Fig.~\ref{fig:fig3}) imply the existence of an asymptotic convergence to a state in which ion mass transfer is dominated by gravitational effects. Fig.~\ref{fig:fig3}a reveals that at a fixed $\Delta \phi=60V_T$, and beyond $Ra\approx 2\times10^6$ the current density becomes insensitive to $\kappa$. In this case we even demonstrate for $\kappa=0$, which signifies a hypothetical case of inactive EKI, the resulting current is close to that of finite $\kappa$. The asymptotic $Ra$ number depends however, depends on $\Delta \phi$ since it sets the base state. For a higher applied potential as in Fig.~\ref{fig:fig3}b ($\Delta \phi=80V_T$), this asymptotic regime is reached at a higher $Ra\approx 1\times10^7$. Our conclusion is that for the flow regimes relevant to electrochemical cells involving aqueous electrolytes, beyond the asymptotic $Ra$ number on the order of $10^6$ to $10^7$, the EKI effects can be safely ignored.

To further investigate the lower $Ra$ regime, in which EKI and RB are strongly coupled, we recast our numerical $\langle I \rangle$ data obtained for the gravitationally unstable orientation with power law fittings in the form of $\langle I \rangle= y_0+mRa^n$ at each $\Delta \phi$.~(Fig.~\ref{fig:fig4}a) The intercept $y_0$ is supposedly the contribution of EKI to the current density, \textit{i.e.} $I^{EKI}|_{Ra=0}=f(\Delta \phi,\kappa)$. Here even for a constant $\kappa=0.5$, the fitting parameters $m$ and $n$ are not constants but instead dependent on $\Delta \phi$ and $\kappa$, opposed to the previous studies on electrochemical RB convection.~\cite{Fenech1960, Ward, Goldstein,Wilke} In these studies $n$ was reported to be a constant, often $\sim 0.3$ in agreement with the power law scalings reported for the thermal RB convection~\cite{Ahlers}. However, these previous correlations on electrochemical RB convection were obtained in the low voltage diffusion limited regime and did not involve the EKI. Consistently our results for $\kappa=0$ indicate that the current shows a good power law correlation $mRa^{0.3}$ for all $\Delta \phi$. (Fig.~\ref{fig:fig4}b) Here the prefactors `$m$' vary with $\Delta \phi$ as the extent of initial density stratification, $\textit{i.e.}$ the extent of ion concentration polarization depend on the electric forcing $\Delta \phi$. 

In the presented $\Delta \phi$, $\kappa$ and $Ra$ number ranges, the EKI and RB instability do not linearly contribute to the total current density; thereby for a given $\Delta \phi$ and $\kappa$, $\langle I \rangle \not= I^{RB}|_{\kappa=0} + I^{EKI}|_{Ra=0}$. Instead we propose the total current to be $\langle I \rangle(\kappa, Ra, \Delta\phi) =  I^{RB}(Ra, \Delta\phi)|_{\kappa=0} + I^* (\kappa, Ra, \Delta\phi)$. Here $I^* \not= I^{EKI}|_{Ra=0}$, but rather an additional current gained from the nonlinear coupling of EKI and RB convection. As seen in Fig.~\ref{fig:fig4}c, $I^*$ reveals a slow exponential decay with respect to $Ra$ number for all $\kappa$. Here we plot the regression lines $I^* = a \times e^{-bRa}$ obtained for $\Delta \phi =80V_T$  for all $\kappa$ where $a=f(\Delta \phi, \kappa)$ and $b=g(\Delta \phi, \kappa)$ are fitting parameters. Our regression analyses for $I^{RB}|_{\kappa=0}$ and $I^*$ show very good correlations for all $\kappa$ and $\Delta \phi$ in the presented $Ra$ number and $\Delta \phi$ ranges where both EKI induced vortices and RB plumes contribute to the total current. We attribute the absence of a universal fitting for `$m$' , `$a$'  and `$b$', to the highly nonlinear nature of the coupling between EKI and RB modes of flow. 

In summary we have analyzed the buoyancy effects on electrokinetic chaos induced at ion selective interfaces over a comprehensive range of parameters for the first time. Our results indicate that in the gravitationally stable scenario, buoyancy effects limit the growth of EKI vortices towards the buoyantly unstable ion-selective interface, while the EKI in the depleted layer remains strong. In the gravitationally unstable scenario, we demonstrated the interplay between two modes of mixing and quantified the asymptotic states in which the transport is dominated by either EKI or RB mechanism. In practical scenarios the quantified thresholds can be translated into system dimensions and experimental salt concentrations; thereby can provide significant guiding insights for the design and scaling analysis of a wide range of electrochemical systems. 
\begin{acknowledgments}
The authors gratefully acknowledge Professor Andreas Acrivos for his invaluable comments on the manuscript and Scott Davidson for providing with the parallelized DNS code. E.K. acknowledges the Rubicon grant from the Netherlands Scientific Organization NWO. 
\end{acknowledgments}
\nocite{*}

\end{document}